\documentclass[final,5p,times, sort&compress, fleqn,twocolumn]{elsarticle}
\journal{Journal of Magnetism and Magnetic Materials}

\usepackage{xcolor}

\usepackage{amsmath}

\begin{document}

\begin{frontmatter}
\title{Insight into ground-state spin arrangement and bipartite entanglement of the polymeric coordination compound [Dy$_2$Cu$_2$]$_n$ through the~symmetric spin-1/2 Ising-Heisenberg orthogonal-dimer chain}
   
\author{Lucia G\'alisov\'a}
\address{Institute of Manufacturing Management, 
  	     Faculty of Manufacturing Technologies with the seat in Pre\v{s}ov, Technical University of~Ko\v{s}ice, \\
  	     Bayerova 1, 080\,01 Pre\v{s}ov, Slovakia}

\ead{lucia.galisova@tuke.sk}

\begin{abstract}
The ground-state spin arrangement and the bipartite entanglement within Cu$^{2+}$\!-\,Cu$^{2+}$ dimers across the magnetization process of the 4f-3d heterometallic coordination polymer [{Dy(hfac)$_2$(CH$_3$OH)}$_2${Cu(dmg)(Hdmg)}$_2$]$_n$ (H$_2$dmg = dimethylglyoxime, Hhfac = 1,1,1,5,5,5-hexafluoropentane-2,4-dione) are theoretically examined using the symmetric isotropic spin-$1/2$ Ising-Heisenberg orthogonal-dimer chain. The numerical results point to five possible ground states of the compound with three different degrees of the quantum entanglement within Cu$^{2+}$\!-\,Cu$^{2+}$. Besides the standard ferrimagnetic and saturated phases without quantum entanglement of Cu$^{2+}$ ions, which are manifested in low-temperature magnetization curve as wide plateaus at the non-saturated magnetization $16.26\mu_{\rm B}$ and at the saturation value $20.82\mu_{\rm B}$, respectively, one also finds an intriguing singlet-like phase with just partial entanglement within Cu$^{2+}$\!-\,Cu$^{2+}$ and two singlet phases with fully entangled Cu$^{2+}$\!-\,Cu$^{2+}$ dimers. The former quantum phase can be identified in the low-temperature magnetization process as very narrow intermediate plateau at the magnetization $9.27\mu_{\rm B}$ per unit cell, while the latter ones as zero magnetization plateau and intermediate plateau at the magnetization $18.54\mu_{\rm B}$. Non-monotonous temperature variations of the concurrence, through which the entanglement within cooper dimers is quantified, point to the possible temporary thermal activation of the entangled states of Cu$^{2+}$\!-\,Cu$^{2+}$ also above non-entangled ferrimagnetic and saturated phases.
\end{abstract}


\end{frontmatter}

\section{Introduction}
\label{sec:1}

Quantum entanglement of particles has currently caught much attention, because it represents a significant resource for quantum computing as well as quantum information processing~\cite{Ami08,Hor09,Mod12}. In particular, frustrated low-dimensional magnetic systems are of great interest, because they have highly degenerate unconventional ground states carrying high degrees of quantum entanglement~\cite{Lhu01,Lac11,Hon04,Mil15,Men16,Zho17}.  

Although real frustrated magnetic materials have three dimensions, many of them can be well described by one-dimensional (1D) quantum spin models if very weak interactions in two spatial dimensions are neglected~\cite{Lac11}. Among various ones, the spin-$1/2$ Heisenberg orthogonal-dimer chain~\cite{Iva97,Rich98} plays a prominent role. This is because it fairly faithfully captures some basic features of the frustrated Shastry-Sutherland lattice~\cite{Sha81} whose structure appears in many quasi two-dimensional magnets~\cite{Smi91,Miy03,Mic06,Sie08,Kim13}; its ground-state phase diagram features a quantum dimerized and symmetry-breaking tetramerized phases~\cite{Iva97,Rich98} and the corresponding magnetization process contains an infinite series of tiny fractional plateaus at $n/(2n + 2)$ of the saturation magnetization ($n$ is a positive integer number), which reflects the ground state with the period $(n + 1)$ of the unit cell~\cite{Kog00,Sch02a,Sch02b}. 

We currently known a series of isostructural 3d-4f heterometallic polymeric complexes [\{Ln(hfac)$_2$(CH$_3$OH)\}$_2$\{Cu(dmg)(Hdmg)\}$_2$]$_n$ (Ln = Gd,~Dy, Tb, Ho, Er, Pr, Nd, Sm, Eu; H$_2$dmg = dimethylglyoxime; Hhfac = 1,1,1,5,5,5-hexafluoropentane-2,4-dione), abbreviated  [Ln$_2$Cu$_2$]$_n$, which involve the orthogonal-dimer chain motif~\cite{Uek05,Uek07,Oka08,Oka09,Oka11}. However, none of them is an experimental realization of the pure spin-$1/2$ Heisenberg orthogonal-dimer chain. It is all the more surprising that the dysprosium-based member [Dy$_2$Cu$_2$]$_n$ of the series of polymeric compounds copies a peculiar structure of the simpler mixed spin-$1/2$ Ising-Heisenberg orthogonal-dimer chain~\cite{Uek07,Oka08}. Indeed, [Dy$_2$Cu$_2$]$_n$ is formed by regularly alternating highly anisotropic dinuclear units of Dy$^{3+}$ ions, which can be theoretically described by the spin-$1/2$ Ising dimers, and roughly orthogonal almost isotropic dinuclear units of Cu$^{2+}$ ions, which can be approximated by the isotropic spin-$1/2$ Heisenberg dimers. Thanks to its rigorous solution, the spin-$1/2$ Ising-Heisenberg orthogonal-dimer chain and its extended versions are an excellent playground for theoretical investigation many intriguing magnetic properties arising from quantum correlations such as anomalous low-temperature thermodynamics~\cite{Ver13,Pau13,Ver14} and enhanced magnetocaloric effect~\cite{Ver13} near critical fields, fractional magnetization plateaus in the magnetization process~\cite{Ver13,Gal21,Str20}, as well as bipartite entanglement within the Heisenberg dimers~\cite{Pau13,Gal21}. Moreover, it provides a fairly satisfactory theoretical fit for experimental magnetization data reported for the real polymeric compound [Dy$_2$Cu$_2$]$_n$ at the temperature $T=0.5$\,K assuming the specific set of model's parameters~\cite{Str20}. 

In the present paper we build on findings in the recent article~\cite{Str20}. Taking into account the values of interaction coupling constants and gyromagnetic factors of spins fitted from magnetization data of [Dy$_2$Cu$_2$]$_n$, we theoretically examine a possible bipartite entanglement within  Cu$^{2+}$\!-\,Cu$^{2+}$ dimers in this polymer through the symmetric version of the isotropic spin-$1/2$ Ising-Heisenberg orthogonal-dimer chain. The phenomenon is examined in detail across the magnetization process as well as upon temperature increase.

The outline of the article is as follows. In Sec.~\ref{sec:2} we briefly describe the spin-$1/2$ Ising-Heisenberg orthogonal-dimer chain used for modelling the coordination polymeric compound [Dy$_2$Cu$_2$]$_n$ and the important steps of its exact analytical treatment leading to accurate numerical results for the sub-lattice and total magnetization, pair correlation functions, and the concurrence are also recalled. The most interesting numerical results are presented in Sec.~\ref{sec:3}. Finally,  Sec.~\ref{sec:4} outlines conclusions and our future outlooks.

\section{Theoretical model of [Dy$_2$Cu$_2$]$_n$ and its solution}
\label{sec:2}

\subsection{Hamiltonian of the model}
\label{subsec:21}

\begin{figure}[t!]
	\centering
	\includegraphics[width=0.95\columnwidth]{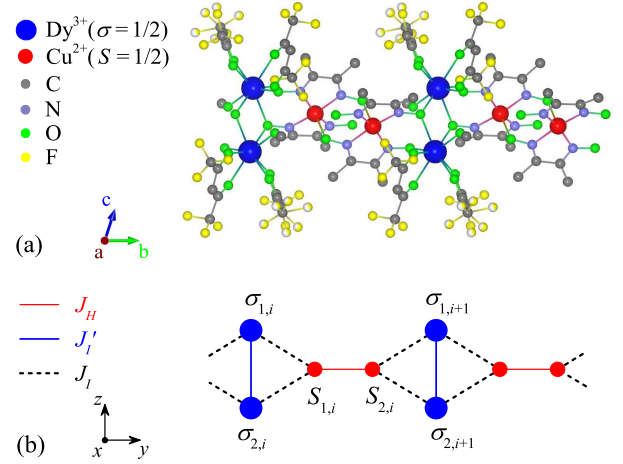}
	\vspace{0mm}
	\caption{(a) The crystal structure of the coordination polymer [Dy$_2$Cu$_2$]$_n$ (full chemical formula is listed in the text) visualized according to crystallographic data from the Crystallography Open Database$^{1}$~\cite{COD}. Hydrogen atoms and corresponding pathways are not shown for clarity. (b) The magnetic structure of the corresponding spin-$1/2$ Ising-Heisenberg orthogonal-dimer chain, in which the Ising spins $\sigma$ correspond to the Dy$^{3+}$ magnetic ions, the Heisenberg spins $S$ correspond to the Cu$^{2+}$ magnetic ions. The coupling constants $J_{H}$, $J_{I}^{\prime}$, and $J_{I}$ match the isotropic Heisenberg intra-dimer interaction between two Dy$^{3+}$ ions, the Ising intra-dimer interaction between two Cu$^{2+}$ ions, and the Ising inter-dimer interaction between Dy$^{3+}$ and Cu$^{2+}$ ions, respectively.
	}
	\label{fig:1}
\end{figure}

The 3d-4f isostructural bimetallic polymeric compound [\{Dy(hfac)$_2$(CH$_3$OH)\}$_2$\{Cu(dmg)(Hdmg)\}$_2$]$_n$, usually referred as [Dy$_2$Cu$_2$]$_n$, crystallizes in a $P\overline{1}$ space group into identical quasi-diamond tetranuclear cells [CuDy$_2$Cu] with the parameters $a = 10.5791(9)\,\AA$, $b = 11.0987(13)\,\AA$, $c = 14.1662(12)\,\AA$, $\alpha = 71.530(4)^{\circ}$, $\beta = 75.669(3)^{\circ}$, $\gamma = 84.702(5)^{\circ}$~\cite{Oka08}. The nearest neighbouring cells are coupled with each other via dinuclear cooper linkages Cu$^{2+}$\!-\,Cu$^{2+}$ realized along the crystallographic $b$-axis. A part of the crystal structure of [Dy$_2$Cu$_2$]$_n$ without unimportant hydrogen atoms and corresponding exchange pathways inside is displayed in Fig.~\ref{fig:1}(a). The visualization was made from crystallographic data accessible in the Crystallography Open Database\footnote{http://www.crystallography.net, COD ID 4001146}~\cite{COD} using the free VESTA software~\cite{VESTA}. 
Here we see 1D orthogonal arrangement of the double oxo-bridged dinuclear ion entities \mbox{Dy$^{3+}$\!-\,Dy$^{3+}$} and macrocyclic dinuclear ones Cu$^{2+}$\!-\,Cu$^{2+}$ which can be quite well described by the symmetric isotropic spin-$1/2$ Ising-Heisenberg orthogonal-dimer chain schematically shown in Fig.~\ref{fig:1}(b). 
It is possible to do so because the  Dy$^{3+}$ ions represent Kramers ions with the ground-state multiplet $^6$H$_{15/2}$ which is a subject of strong crystal-field splitting into eight well separated Kramers doublets~\cite{Jon74,Jen91}, even though two different types of the oximate O--N bridges are realized between Cu$^{2+}$ and Dy$^{3+}$ ions in the real [CuDy$_2$Cu] cell~\cite{Oka08}. Our recent study~\cite{Str20}, providing a satisfactory theoretical fit of the magnetization data reported for [Dy$_2$Cu$_2$]$_n$ at the temperature $T=0.5$\,K, is the convincing evidence for this.

Returning to the crystal structure of the polymeric compound [Dy$_2$Cu$_2$]$_n$ shown in Fig.~\ref{fig:1}(a), the dysprosium dimers \mbox{Dy$^{3+}$\!-\,Dy$^{3+}$} along the crystallographic $c$-axis can be treated as the standard Ising spin pairs with the magnitude $1/2$, while the cooper ones Cu$^{2+}$\!-\,Cu$^{2+}$ along $b$-axis can be regarded as the isotropic spin-$1/2$ Heisenberg dimers.
Moreover, if oximate bridges between adjacent Cu$^{2+}$ and Dy$^{3+}$ ions are approximated by the Ising interaction of the same strength, the total Hamiltonian of the spin-$1/2$ Ising-Heisenberg orthogonal-dimer chain for capturing magnetic features of the polymer [Dy$_2$Cu$_2$]$_n$ can be written as [see Fig.~\ref{fig:1}(b)]:
\begin{eqnarray}
\label{eq:H1}
\hat{H} \!\!\!\!&=&\!\!\!\! J_{I}^{\prime}\sum_{i = 1}^{N}\hat{\sigma}_{1,i}^{z}\hat{\sigma}_{2,i}^{z} + 
J_{H}\sum_{i = 1}^{N}\hat{\bf S}_{1,i}\cdot\hat{\bf S}_{2,i}
\nonumber
\\
\!\!\!\!&+&\!\!\!\!
J_{I}\sum_{i = 1}^{N}\left[\hat{S}_{1,i}^{z}(\hat{\sigma}_{1,i}^{z}+\hat{\sigma}_{2,i}^{z})+\hat{S}_{2,i}^{z}(\hat{\sigma}_{1,i+1}^{z}+\hat{\sigma}_{2,i+1}^{z})\right] 
\nonumber
\\
\!\!\!\!&-&\!\!\!\!
g_{I}\mu_{\rm B}B\sum_{i = 1}^{N}(\hat{\sigma}_{1,i}^{z}+\hat{\sigma}_{2,i}^{z}) - g_{H}\mu_{\rm B}B\sum_{i = 1}^{N}(\hat{S}_{1,i}^{z} + \hat{S}_{2,i}^{z})\,.
\end{eqnarray}
In above, $\hat{\sigma}_{(1)2,i}^{z}$, $\hat{\sigma}_{(1)2,i+1}^{z}$ label $z$-components of the spin-$1/2$ operator assigned to Dy$^{3+}$ ions, while $\hat{\bf S}_{1(2),i}$ labels the tree-component spin-$1/2$ operator of the Heisenberg spins assigned to Cu$^{2+}$ magnetic ions. The coupling constants $J_{I}^{\prime}$ and  $J_{H}$ account for the Ising and isotropic Heisenberg intra-dimer interactions within Dy$^{3+}$\!-\,Dy$^{3+}$ dimers and to them almost orthogonal Cu$^{2+}$\!-\,Cu$^{2+}$ dimers, respectively. The term $J_{I}$ in second line of the Hamiltonian~(\ref{eq:H1}) determines a strength of the Ising inter-dimer interaction between the nearest Cu$^{2+}$ and Dy$^{3+}$ ions. The last two terms $g_{I}\mu_{\rm B}B$, $g_{H}\mu_{\rm B}B$ in third line of Eq.~(\ref{eq:H1}), which differ from each other only by the gyromagnetic factors $g_{I}$, $g_{H}$ of corresponding spins ($\mu_{\rm B}$ is the Bohr magneton), represent the standard Zeeman's terms taking into account different magnetostatic energies for magnetic moments of the Ising and Heisenberg spins in the external magnetic field $B$ measured in, e.g., Teslas, respectively. In the present model, the magnetic field $B$ is applied along the $z$-axis, which approximates the crystallographic $c$-axis in the real polymer [Dy$_2$Cu$_2$]$_n$ ($z\sim c$, $g_{H} \sim g_{\rm Dy}^{c}$, see Fig.~\ref{fig:1}). 
Finally, $N$ marks the total number of the Ising and Heisenberg dimers in the model and the periodic boundary condition $\hat{\sigma}_{1(2),N+1}^{z} \equiv \hat{\sigma}_{1(2),1}^{z}$ is assumed to eliminate boundary effect.

Due to the specific distribution of the exchange interactions in the considered theoretical model, the total Hamiltonian~(\ref{eq:H1}) can be alternatively expressed in the symmetric form, namely as the sum of $N$ identical commuting six-spin cluster Hamiltonians $\hat{H} = \sum_{i = 1}^{N} \hat{H}_i$,
where each $\hat{H}_i$ involves all interaction and field terms related to the respective horizontal Heisenberg dimer (or, equivalently, the Cu$^{2+}$\!-\,Cu$^{2+}$ dimer) and two adjacent vertical Ising (Dy$^{3+}$\!-\,Dy$^{3+}$) dimers:
\begin{eqnarray}
\label{eq:Hi}
\hat{H}_i \!\!\!\!&=&\!\!\! 
\frac{J_{I}^{\prime}}{2}(\hat{\sigma}_{1,i}^{z}\hat{\sigma}_{2,i}^{z}+ \hat{\sigma}_{1,i+1}^{z}\hat{\sigma}_{2,i+1}^{z}) + J_{H}\hat{\bf S}_{1,i}\cdot\hat{\bf S}_{2,i}
\nonumber
\\
\!\!\!\!&+&\!\!\!\!
J_{I}\left[\hat{S}_{1,i}^{z}(\hat{\sigma}_{1,i}^{z}+\hat{\sigma}_{2,i}^{z})+\hat{S}_{2,i}^{z}(\hat{\sigma}_{1,i+1}^{z}+\hat{\sigma}_{2,i+1}^{z})\right] \nonumber
\\
\!\!\!\!&-&\!\!\!\!
\frac{g_{I}\mu_{\rm B}B}{2}(\hat{\sigma}_{1,i}^{z}+\hat{\sigma}_{2,i}^{z}+\hat{\sigma}_{1,i+1}^{z}+\hat{\sigma}_{2,i+1}^{z}) \nonumber
\\
\!\!\!\!&-&\!\!\!\!
g_{H}\mu_{\rm B}B(\hat{S}_{1,i}^{z} + \hat{S}_{2,i}^{z})\,. 
\end{eqnarray}
The factor $1/2$ at the Ising intra-dimer coupling constant $J_{I}^{\prime}$ and the Zeeman's term $g_{I}\mu_{\rm B}B$ prevents a double counting of these two interaction terms into the total Hamiltonian's expression.

The eigenvalues of the Hamiltonian~(\ref{eq:Hi}) can be obtained in the local orthonormal basis $\big\{|S_{1,i}^{z}, S_{2,i}^{z}\rangle\big\}=\big\{|\!\uparrow\uparrow\rangle_{i}, |\!\uparrow\downarrow\rangle_{i},$ $|\!\downarrow\uparrow\rangle_{i}, |\!\downarrow\downarrow\rangle_{i}\big\}$ of the $i$th Heisenberg pair corresponding to the  Cu$^{2+}$\!-\,Cu$^{2+}$ dimer. Their explicit expressions contain all interaction and field terms present in $\hat{H}_i$, as well as the eigenvalues $\sigma_{T,i(i+1)}^{z} = \{-1,0,1\}$ of the total spin operators $\hat{\sigma}_{T,i(i+1)}^{z} = \hat{\sigma}_{1,i(i+1)}^{z}+\hat{\sigma}_{2,i(i+1)}^{z}$ corresponding to the $i$th and $(i+\!1)$st Ising dimers:
\begin{subequations}
\begin{eqnarray}
\label{eq:eigenvalue1}
E_{i,1}(\sigma_{T,i}^{z},\sigma_{T,i+1}^{z})\!\!\!\!&=&\!\!\!\!
\,\frac{J_{I}^{\prime}}{4}\left[\big(\sigma_{T,i}^{z}\big)^2+\big(\sigma_{T,i+1}^{z}\big)^2-1\right] +\frac{J_{H}}{4} 
\nonumber\\ 
\!\!\!\!&-& \!\!\!\!
\frac{J_{I}+g_{I}\mu_{\rm B}B}{2}\big(\sigma_{T,i}^{z}+\sigma_{T,i+1}^{z}\big) +
g_{H}\mu_{\rm B}B,
\hspace{0.75cm} \\
\label{eq:eigenvalue2}
E_{i,2}(\sigma_{T,i}^{z},\sigma_{T,i+1}^{z})\!\!\!\!&=&\!\!\!\!
\,
\frac{J_{I}^{\prime}}{4}\left[\big(\sigma_{T,i}^{z}\big)^2+\big(\sigma_{T,i+1}^{z}\big)^2-1\right] + \frac{J_{H}}{4}  \nonumber\\ 
\!\!\!\!&+&\!\!\!\! 
\frac{J_{I} - g_{I}\mu_{\rm B}B}{2}\big(\sigma_{T,i}^{z}+\sigma_{T,i+1}^{z}\big) -
g_{H}\mu_{\rm B}B,
\\
\label{eq:eigenvalue3}
E_{i,3}(\sigma_{T,i}^{z},\sigma_{T,i+1}^{z})\!\!\!\!&=&\!\!\!\!
\frac{J_{I}^{\prime}}{4}\left[\big(\sigma_{T,i}^{z}\big)^2+\big(\sigma_{T,i+1}^{z}\big)^2-1\right] -\frac{J_{H}}{4}  
\nonumber \\
\!\!\!\!&+&\!\!\!\!
\frac{1}{2}\!\sqrt{J_{I}^{2}\big(\sigma_{T,i}^{z}-\sigma_{T,i+1}^{z}\big)^{2}+J_{H}^2} 
\nonumber\\ 
\!\!\!\!&-& \!\!\!\!
\frac{g_{I}\mu_{\rm B}B}{2}\big(\sigma_{T,i}^{z}+\sigma_{T,i+1}^{z}\big),
\\
\label{eq:eigenvalue4}
E_{i,4}(\sigma_{T,i}^{z},\sigma_{T,i+1}^{z})\!\!\!\!&=&\!\!\!\!
\frac{J_{I}^{\prime}}{4}\left[\big(\sigma_{T,i}^{z}\big)^2+\big(\sigma_{T,i+1}^{z}\big)^2-1\right]  -\frac{J_{H}}{4}  
\nonumber\\ 
\!\!\!\!&-& \!\!\!\!
\frac{1}{2}\!\sqrt{J_{I}^{2}\big(\sigma_{T,i}^{z}-\sigma_{T,i+1}^{z}\big)^{2}+J_{H}^{2}}
\nonumber\\ 
\!\!\!\!&-& \!\!\!\!
\frac{g_{I}\mu_{\rm B}B}{2}\big(\sigma_{T,i}^{z}+\sigma_{T,i+1}^{z}\big).
\end{eqnarray}
\end{subequations}
The respective eigenvectors are:
\begin{subequations}
\label{eq:eigenvectors}
\begin{eqnarray}
\label{eq:eigenvector1}
|\psi\rangle_{i,1} \!\!\!\!&=&\!\!\!\! |\!\downarrow\downarrow\rangle_{i},
\\
\label{eq:eigenvector2}
|\psi\rangle_{i,2} \!\!\!\!&=&\!\!\!\! |\!\uparrow\uparrow\rangle_{i},
\\
\label{eq:eigenvector3}
|\psi\rangle_{i,3} \!\!\!\!&=&\!\!\!\! \sin\varphi|\!\uparrow\downarrow\rangle_{i} + \cos\varphi|\!\downarrow\uparrow\rangle_{i},  
\\
\label{eq:eigenvector4}
|\psi\rangle_{i,4} \!\!\!\!&=&\!\!\!\! \sin\varphi|\!\uparrow\downarrow\rangle_{i} - \cos\varphi|\!\downarrow\uparrow\rangle_{i}.
\end{eqnarray}
\end{subequations}
Evidently, the former two eigenvectors~(\ref{eq:eigenvector1}) and (\ref{eq:eigenvector2}) describe standard ferromagnetic spin arrangements of the $i$th Heisenberg dimer with $z$-projections  $S_{T,i}^z = -1$ and $1$ of its total spin, respectively. 
The latter ones~(\ref{eq:eigenvector3}) and (\ref{eq:eigenvector4}) correspond to remarkable quantum states of two admissible antiferromagnetic states with $S_{T,i}^z = 0$. The mixing angle $\varphi$, determining probability amplitudes of these states, is defined through the formula $\tan(2\varphi) = J_{H}/\big[J_{I}\big(\sigma_{T,i}^{z}-\sigma_{T,i+1}^{z}\big)\big]$.

Referring to the most satisfactory theoretical fit of the experimental magnetization data for [Dy$_2$Cu$_2$]$_n$ discussed in the recent article~\cite{Str20}, the parameter set occurring in the cluster Hamiltonian~(\ref{eq:Hi}) and its eigenvalues~(\ref{eq:eigenvalue1})--(\ref{eq:eigenvalue4}) will be hereafter fixed to the following specific values: $J_{H}/k_{\rm B} = 1.73$\,K, $J_{I}^\prime/k_{\rm B} = 17.35$\,K, $J_{I}/k_{\rm B} = 8.02$\,K,  $g_{H} = 2.28$, $g_{I}= 18.54$. It should be noted that the first value for the Heisenberg intra-dimer interaction agrees with the actual  exchange coupling $J_{\rm Cu-Cu}/k_{\rm B} = 1.73$\,K. Other two for the Ising intra- and inter-dimer constants fall into adequate ranges for the actual value $J_{\rm Dy-Dy}/k_{\rm B} = 0.08$\,K and the mean value $J_{\rm Dy-Cu}/k_{\rm B} = 0.48$\,K (calculated from $J_A/k_{\rm B} = 0.895(8)$\,K, $J_B/k_{\rm B} = 0.061(8)$\,K for two different exchange pathways between Dy$^{3+}$ and Cu$^{2+}$ ions in [CuDy$_2$Cu] using the effective Ising tetranuclear model~\cite{Oka08}) after dividing by the factors $225$ and $15$, respectively.\footnote{The dividing factors follow from the true value of the total angular momentum  $J_{\rm Dy}=15/2$ of the Dy$^{3+}$ ion instead of $\sigma^z=1/2$.} Last but not least, the estimated gyromagnetic factors can also be accepted; the former one for the Heisenberg spins is in a good agreement with the typical value $g_{\rm Cu}^{z}=2.2$ for Cu$^{2+}$ ions, while the latter, corresponding to the Ising spins, is only $7.3\%$ lower than the theoretically calculated value $g_{\rm Dy}^{z}=20$ for the effective Ising-spin picture of the Dy$^{3+}$ ion provided that $B\parallel z$~\cite{Jon74}.

\subsection{Partition function, Gibbs free energy, magnetization and~pair correlation functions}
\label{subsec:22}

Complete rigorous treatment of the spin-$1/2$ Ising-Heisenberg orthogonal-dimer chain was reported for its more general anisotropic version in our preceding article~\cite{Gal21}. Specifically, the partition function of the model, Gibbs free energy per unit cell, as well as local sub-lattice magnetization per Ising and Heisenberg spin, total magnetization per spin, and pair correlation functions of the nearest-neighbouring spins have been derived using the well-known transfer-matrix formalism~\cite{Bax82}.

To recall the main idea of the mentioned algebraic approach, its using allows one to express the partition function of the isotropic spin-$1/2$ Ising-Heisenberg orthogonal-dimer chain given by the Hamiltonian~(\ref{eq:H1}) as a trace of $N$th power of the four-by-four transfer matrix ${\rm\bf T}$ related to the cluster Hamiltonian~(\ref{eq:Hi}): 
\begin{eqnarray}
\label{eq:Z}
Z \!\!\!&=&\!\!\!\! \textrm{Tr}\,{\rm e}^{-\hat{H}/(k_{\rm B}T)} = \sum_{\{\sigma_{1,i},\sigma_{2,i}\}}\prod_{i=1}^{N}\textrm{Tr}_{S_{1,i},S_{2,i}}\,{\rm e}^{-\hat{H}_i/(k_{\rm B}T)}  
\nonumber \\
\!\!\!&=&\!\!\!\!\! \sum_{\{\sigma_{1,i},\sigma_{2,i}\}}\prod_{i=1}^{N} {\rm  T}(\sigma_{1,i}^{z}, \sigma_{2,i}^{z};\sigma_{1,i+1}^{z}, \sigma_{2,i+1}^{z}) = \textrm{Tr}\,{\rm\bf T}^{N}.
\end{eqnarray}
In above, $k_{\rm B}$ is the Boltzmann's constant, $T$ is the temperature of the system, the summation $\sum_{\{\sigma_{1,i},\sigma_{2,i}\}}$ is carried out over all possible configurations of the vertical Ising dimers, the product $\prod_{i=1}^{N}$ gradually runs over all six-spin clusters, and $\textrm{Tr}_{S_{1,i},S_{2,i}}$ stands for a trace over all degrees of freedom of the $i$th horizontal Heisenberg dimer. The elements of the matrix ${\rm\bf T}$ are given by four eigenvalues~(\ref{eq:eigenvalue1})--(\ref{eq:eigenvalue4}) of the Hamiltonian~(\ref{eq:Hi}) through the formula: 
\begin{equation}
\label{eq:T}
{\rm T}(\sigma_{1,i}^{z}, \sigma_{2,i}^{z};\sigma_{1,i+1}^{z}, \sigma_{2,i+1}^{z})  =  \textrm{Tr}_{S_{1,i},S_{2,i}}\,{\rm e}^{-\hat{H}_i/(k_{\rm B}T)} = \sum_{j=1}^{4}{\rm e}^{-E_{i,j}/(k_{\rm B}T)}.
\end{equation}
In the thermodynamic limit $N\to \infty$, the partition function~(\ref{eq:Z}) of the considered chain as well as its Gibbs free energy are determined by the largest of four eigenvalues of the transfer matrix~${\rm\bf T}$. These can be obtained using the standard algebra. More computational details and explicit expressions of all four eigenvalues of ${\rm\bf T}$ can be taken from Ref.~\cite{Gal21}, just set $\Delta=1$ in parameters below Eq.~(8b).  

The Gibbs free energy per six-spin cluster of the investigated theoretical model given by:
\begin{equation}
\label{eq:G}
G = -k_{\rm B}T\lim_{N\to \infty}\frac{1}{N}\ln Z
\end{equation}
subsequently allows to calculate the total magnetization $m$ per elementary unit cell of the polymeric compound [Dy$_2$Cu$_2$]$_n$ using the well-known relation:
\begin{equation}
\label{eq:m}
m =  -\frac{\partial G}{\partial B}.
\end{equation} 

To gain a deeper insight into the magnetization process, it is also useful to calculate the local Ising and Heisenberg sub-lattice magnetization normalized per, e.g., elementary unit of the proposed model. These are determined by a product of the corresponding $g$-factor, Bohr magneton, and average of $z$-component of the appropriate spin operator:
\begin{equation}
\label{eq:mImH}
m_{I} = g_{I}\mu_{\rm B}\langle\hat{\sigma}_{1,i}^{z}+\hat{\sigma}_{2,i}^{z}\rangle,
\quad
m_{H} = g_{H}\mu_{\rm B}\langle\hat{S}_{1,i}^{z}+\hat{S}_{2,i}^{z}\rangle.
\end{equation}
In the established notation, $m_{I}$ corresponds to the Dy$^{3+}$\!-\,Dy$^{3+}$ (Ising) dimer and $m_{H}$ to the Cu$^{2+}$\!-\,Cu$^{2+}$ (Heisenberg) dimer. Rigorous solutions for averages of $z$-component of the Ising and Heisenberg spin operators in Eq.~(\ref{eq:mImH}) can be obtained either as negative derivatives of the Gibbs free energy per unit cell with respect to the Zeeman's terms corresponding to these spins, or by employing the generalized Callen-Suzuki spin identity~\cite{Cal63,Suz65,Bal02} in combination with exact mapping theorems developed by J.H. Barry and co-workers~\cite{Bar88,Kha90,Bar95}.

It should be mentioned that using analogous approach(es), we can also calculate other important physical quantities, namely, the spatial components of the pair correlation functions $\langle\hat{\sigma}_{1,i}^{z}\hat{\sigma}_{2,i}^{z}\rangle$, $\langle\hat{S}_{1,i}^{z}\hat{S}_{2,i}^{z}\rangle$, $\langle\hat{S}_{1,i}^{x(y)}\hat{S}_{2,i}^{x(y)}\rangle$, $\langle\hat{\sigma}_{1(2),i}^{z}\hat{S}_{1,i}^{z}\rangle = \langle \hat{\sigma}_{1(2),i+1}^{z}\hat{S}_{2,i}^{z}\rangle$ of the nearest-neighbouring spins. These will be of great help to understand evolution of mutual correlations between the nearest cooper and dysprosium ionic neighbours in [Dy$_2$Cu$_2$]$_n$ across the magnetization process and with increasing temperature.

\subsection{Concurrence}
\label{subsec:23}

The subject of main interest of the present study is the bipartite entanglement between two exchange-coupled Cu$^{2+}$ magnetic ions forming the horizontal dimers in the polymeric compound [Dy$_2$Cu$_2$]$_n$. The phenomenon can be quantified using the concurrence formalism~\cite{Woo98}. L. Amico and coworkers have shown that the quantity concurrence can be quite elegantly calculated from the local single-site and pair correlation functions~\cite{Ami04,Ost13}. In the established formalism, the concurrence quantifying the entanglement within the horizontal Heisenberg dimers in the considered spin-$1/2$ Ising-Heisenberg orthogonal-dimer chain follows from the relation:
\begin{eqnarray}
\label{eq:C}
C \!\!\!\!&=&\!\!\!\! \max\Bigg\{0, 4\big|\langle \hat{S}_{1,i}^{x(y)}\hat{S}_{2,i}^{x(y)}\rangle\big| {} \nonumber\\
&&\hspace{0.55cm}{}- 2\sqrt{\bigg(\frac{1}{4}+\big|\langle \hat{S}_{1,i}^{z}\hat{S}_{2,i}^{z}\rangle\big|\bigg)^{\!2}\!-\bigg(\frac{1}{2}\langle\hat{S}_{1,i}^{z}+\hat{S}_{2,i}^{z}\rangle\bigg)^{\!2}}\,\Bigg\}.\hspace{0.75cm}
\end{eqnarray}

\section{Numerical results}
\label{sec:3}

In the current section, we proceed to a presentation of the rigorous numerical results obtained for the isotropic spin-$1/2$ Ising-Heisenberg orthogonal-dimer chain proposed in preceding subsection with the aim to theoretically approximate the ground state, magnetization process, and the quantum and  thermal bipartite entanglement within Cu$^{2+}$\!-\,Cu$^{2+}$  dimers of the polymeric compound [Dy$_2$Cu$_2$]$_n$. 

\subsection{Ground state}
\label{subsec:31}

\begin{figure}[t!]
	\centering
	\vspace{0mm}
	\includegraphics[width=1.0\columnwidth]{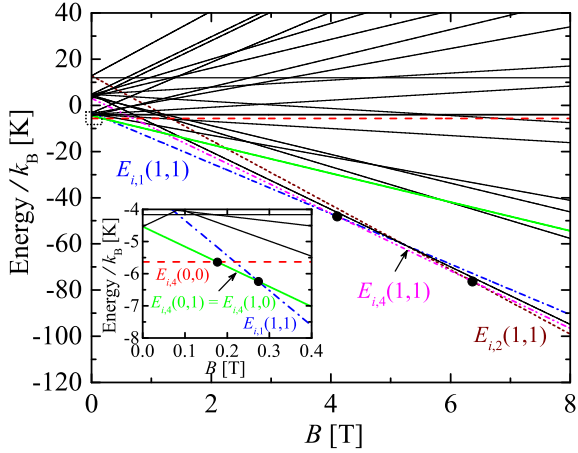}
	\vspace{-6mm}
	\caption{The complete energy spectrum of the cluster Hamiltonian~(\ref{eq:Hi}) versus magnetic field $B$. The inset shows the narrow region of very low magnetic fields marked by black dotted rectangle in the enlarged scale. The colored thicker lines of different styles represent energy levels determining the ground state spin arrangement in a given field range. The black circles highlight intersections of these levels at specific critical fields.}
	\label{fig:2}
\end{figure}

We start with the analysis of possible ground-state spin arrangement of [Dy$_2$Cu$_2$]$_n$. For this purpose, the complete energy spectrum of the cluster Hamiltonian~(\ref{eq:Hi}) versus external magnetic field $B$ is depicted in Fig.~\ref{fig:2}. 
It was obtained from the eigenvalues~(\ref{eq:eigenvalue1})--(\ref{eq:eigenvalue4}) by setting the exchange integrals $J_{H}$, $J_{I}^{\prime}$, $J_{I}$ and $g$-factors $g_{H}$, $g_{I}$ to the values specified at the end of Subsec.~\ref{subsec:21} and by considering all admissible combinations of the quantum numbers $\sigma_{T,i(i+1)}^{z} = \{-1,0,1\}$ for the total Ising spin operators $\hat{\sigma}_{T,i(i+1)}^{z}$. The values of plotted $E_{i,j}$ are normalized to the Boltzmann's constant $k_{\rm B}$ and expressed in Kelvins.

It is obvious form Fig.~\ref{fig:2} that the external magnetic field~$B$ induces in total five different lowest-energy levels, which clearly indicates five possible ground-state spin arrangements of [Dy$_2$Cu$_2$]$_n$. Namely, at zero and very low magnetic fields $B<0.117$\,T, the lowest energy level is $E_{i,4}(0,0)$. It corresponds to the four-fold degenerate configuration of the $i$th six-spin cluster, where the horizontal Heisenberg dimer occupies a perfect singlet state given by the eigenvector~(\ref{eq:eigenvector4}) with $\varphi = 45^{\circ}$ and the surrounding vertical Ising dimers have zero total spins due to two-fold degenerate antiferromagnetic spin arrangements. On the other hand, if $B$ is higher than $0.117$\,T but lower than $0.274$\,T, then, two equivalent two-fold degenerate energy levels $E_{i,4}(1,0) = E_{i,4}(0,1)$ can be identified as the lowest ones. Here, the Heisenberg dimer still remains in a quantum superposition of two antiferromagnetic microstates $|\!\uparrow\downarrow\rangle_{i}$ and $|\!\downarrow\uparrow\rangle_{i}$ given by Eq.~(\ref{eq:eigenvector4}), but the probability amplitudes of these microstates differ from each other due to the mixing angle, which is $\varphi\approx6^{\circ}$ for $E_{i,4}(1,0)$ and $\varphi\approx354^{\circ}$ for $E_{i,4}(0,1)$. The surrounding vertical Ising dimers are ferromagnetically and antiferromagnetically ordered, as indicated by the respective total spin values $\sigma_{T,i(i+1)}^{z} = 1$ and $\sigma_{T,i+1(i)}^{z} =0$ in the energy labels. For the magnetic fields $B>0.274$\,T, three more lowest energy levels can be found, namely, $E_{i,1}(1,1)$ in the range $(0.274-4.107)$\,T, $E_{i,4}(1,1)$ from $4.107$\,T to $6.366$\,T, and, $E_{i,2}(1,1)$ above the value $6.366$\,T. Clearly, the listed levels correspond exclusively to cluster configurations with maximum total spins $\sigma_{T,i}^{z} = \sigma_{T,i+1}^{z} = 1$, which implies a permanent full polarization of both outer Ising dimers. A more complex situation can be found in the case of the Heisenberg dimer. According to the respective eigenvectors~(\ref{eq:eigenvector1}), (\ref{eq:eigenvector2}), the energy levels $E_{i,1}(1,1)$, $E_{i,2}(1,1)$ describe a unique ferromagnetic spin ordering of the Heisenberg pair with the total spin projections $S_{T,i}^z = -1$, $1$, respectively. The negative projection $S_{T,i}^z = -1$ points out to the opposite spin alignment with respect to the fields $B\in (0.274-4.107)$\,T, while the positive one $S_{T,i}^z = 1$ proves the full polarization of the Heisenberg dimer to the field direction after exceeding the saturation value $B_{sat} = 6.366$\,T. Last but not least, the level $E_{i,4}(1,1)$, located between two preceding ones at magnetic fields $B\in (4.107-6.366)$\,T, is linked to eigenvector~(\ref{eq:eigenvector4}) with the specific value of the mixing angle $\varphi_i = 45^{\circ}$. It thus reflects a perfect singlet state of the Heisenberg dimer with the zero total spin projection $S_{T,i}^z = 0$. 

\begin{figure}[t!]
	\centering
	\vspace{0mm}
	\includegraphics[width=1.0\columnwidth]{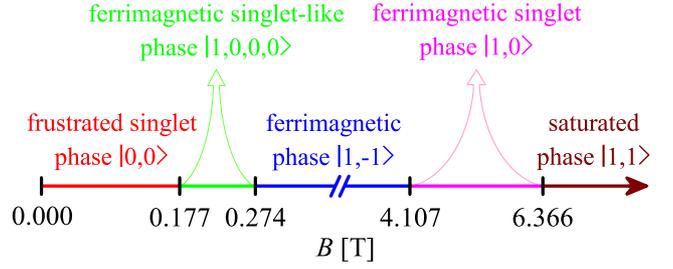}
	\vspace{-6mm}
	\caption{The theoretically predicted ground-state phase diagram for [Cu$_2$Dy$_2$]$_n$ in the external magnetic field $B$ using the spin-$1/2$ Ising-Heisenberg orthogonal-dimer chain with the specific parameter set: $J_{H}/k_{\rm B} = 1.73$\,K, \mbox{$J_{I}^\prime/k_{\rm B} = 17.35$\,K}, $J_{I}/k_{\rm B} = 8.02$\,K, $g_{H} = 2.28$, $g_{I}= 18.54$. The numbers in ket brackets determine $z$-projections of the total spins of the vertical Ising dimer and the neighbouring horizontal Heisenberg dimer, respectively. The label of the ferrimagnetic singlet-like phase includes four numbers due to breaking the translational symmetry of the system.}
	\label{fig:3}
\end{figure}

By extending the findings on the lowest-energy eigenstates of the cluster Hamiltonian (\ref{eq:Hi}) to the whole spin-$1/2$ Ising-Heisenberg orthogonal-dimer chain, it is possible to construct the ground-state phase diagram, which provides a theoretical insight into development of the zero-temperature magnetic ordering in the coordination compound [Dy$_2$Cu$_2$]$_n$ as the external magnetic field $B$ increases. The corresponding phase diagram is schematically depicted in Fig.~\ref{fig:3}. We can see from the figure that it has the form of horizontal magnetic-field axis divided into five parts of various lengths, which are associated with five different phases. The initial part from $0$\,T to $0.117$\,T declares the stability region of the quantum frustrated singlet phase $|0,0\rangle$, in which two-fold degenerate antiferromagnetic arrangements of the vertical Ising dimers regularly alternate with perfect singlet-dimer states of the horizontal Heisenberg pairs:
\begin{eqnarray}
\label{eq:0,0}
|0,0\rangle = \prod_{i=1}^{N}\left.\big|\mbox{\normalsize${\uparrow\atop\downarrow}$}\right\rangle_{\!i} \,\left(\textrm{or}\left.\big|\mbox{\normalsize${\downarrow\atop\uparrow}$}\right\rangle_{\!i}\right)\otimes
\frac{1}{\sqrt{2}}\,
\Big(
|\!\uparrow\downarrow\rangle_{i} - |\!\downarrow\uparrow\rangle_{i}
\Big).
\end{eqnarray}
The following very narrow part of the axis $(0.117-0.274)$\,T corresponds to the remarkable highly degenerate quantum state with a two-fold breaking of translation symmetry, which can be marked as the ferrimagnetic singlet-like phase $|1,0,0,0\rangle$ due to alternating ferro- and antiferromagnetic spin arrangements of the Ising dimers and singlet-dimer-like states of the Heisenberg pairs:
\begin{eqnarray}	
\label{eq:1,0,0,0}
|1,0,0,0\rangle \!\!\!\!&=&\!\!\!\! \prod_{i=1}^{N/2}
\left.\big|\mbox{\normalsize${\uparrow\atop\uparrow}$}\right\rangle_{\!2i-1}\otimes\Big(
\sin 6^{\circ}|\!\uparrow\downarrow\rangle_{2i-1} - \cos 6^{\circ}|\!\downarrow\uparrow\rangle_{2i-1}
\Big) \nonumber\\
&\otimes&\!\!\!\!
\left.\big|\mbox{\normalsize${\uparrow\atop\downarrow}$}\right\rangle_{\!2i} 
\,\left(\textrm{or}\left.\big|\mbox{\normalsize${\downarrow\atop\uparrow}$}\right\rangle_{\!2i}\right)\otimes\Big(
\cos 6^{\circ}|\!\uparrow\downarrow\rangle_{2i} - \sin 6^{\circ}|\!\downarrow\uparrow\rangle_{2i}
\Big).\hspace{0.65cm}
\end{eqnarray}
The remaining parts of the 1D ground-state phase diagram define stability regions of the phases with fully polarized Ising spins. Specifically, in the relatively wide magnetic-field range $(0.274-4.107)$\,T one finds the standard ferrimagnetic phase $|1,-1\rangle$ with a perfect antiparallel arrangement of the neighbouring Ising and Heisenberg dimers:
\begin{eqnarray}
\label{eq:1,-1}
|1,-1\rangle = \prod_{i=1}^{N}
\left.\big|\mbox{\normalsize${\uparrow\atop\uparrow}$}\right\rangle_{\!i}\otimes
|\!\downarrow\downarrow\rangle_{i}\,.
\end{eqnarray}
On the other hand, next range between $4.107$\,T and $6.366$\,T corresponds to the quantum ferrimagnetic singlet phase $|1,0\rangle$, where the fully polarized Ising dimers regularly alternate with perfect singlet-dimer states of the Heisenberg pairs:
\begin{eqnarray}	
\label{eq:1,0}
|1,0\rangle = \prod_{i=1}^{N}
\left.\big|\mbox{\normalsize${\uparrow\atop\uparrow}$}\right\rangle_{\!i}\otimes
\frac{1}{\sqrt{2}}\,\Big(|\!\uparrow\downarrow\rangle_{i} - |\!\downarrow\uparrow\rangle_{i}\Big).
\end{eqnarray}
Finally, the last high-field part $B>6.366$\,T represents the stability region of the fully saturated phase $|1,1\rangle$ with all Ising and Heisenberg spins polarized into magnetic-field direction:
\begin{eqnarray}
\label{eq:1,1,1}
|1,1\rangle = \prod_{i=1}^{N}
\left.\big|\mbox{\normalsize${\uparrow\atop\uparrow}$}\right\rangle_{\!i}\otimes
|\!\uparrow\uparrow\rangle_{i}\,.
\end{eqnarray}

\subsection{Magnetization and correlation functions}
\label{subsec:32}

\begin{figure}[t!]
	\centering
	\vspace{0mm}
	\includegraphics[width=1.0\columnwidth]{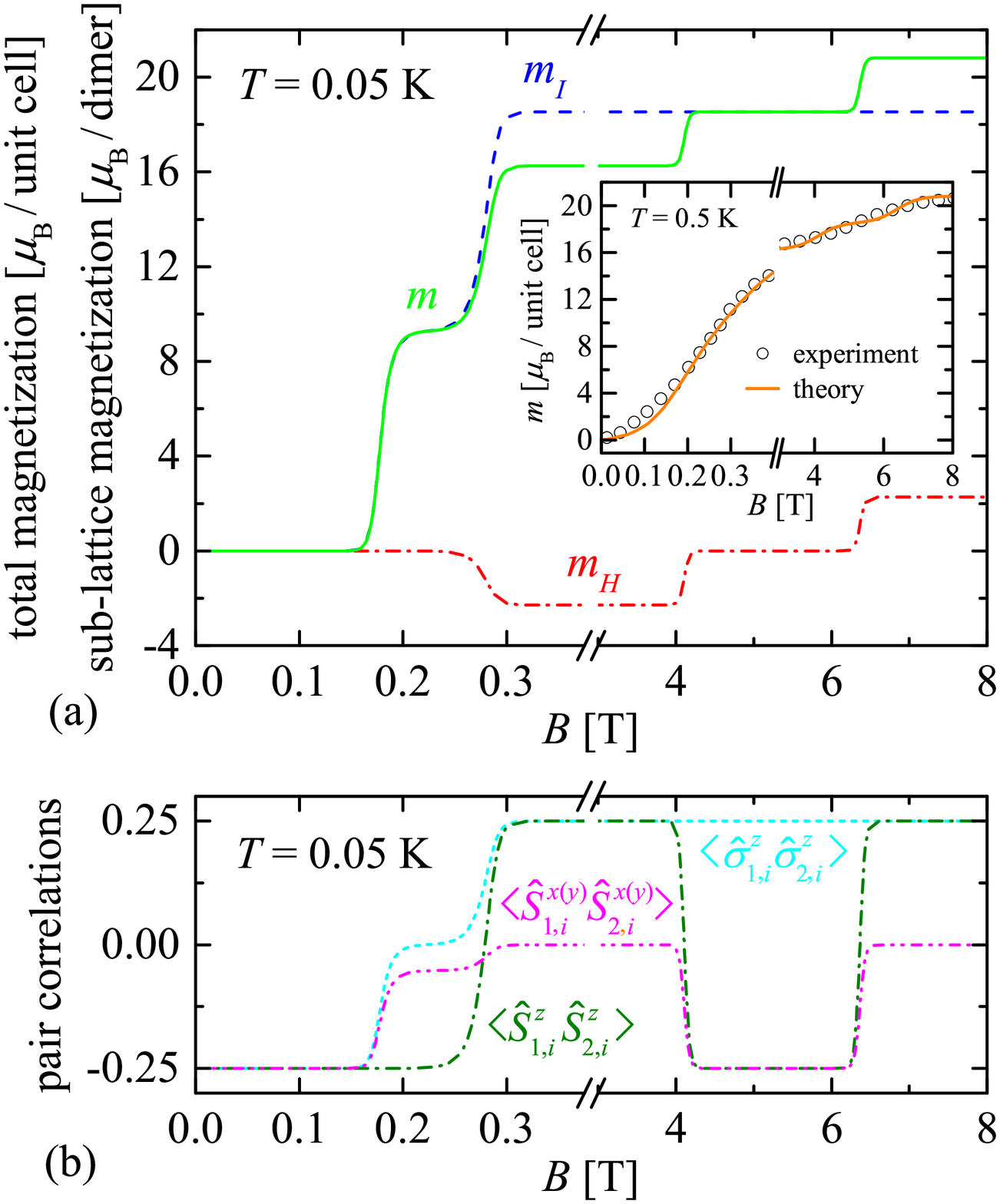}
	\vspace{-7mm}
	\caption{The theoretical plots of (a) the sub-lattice magnetization $m_{I}$ and $m_{H}$ per Ising and Heisenberg spin dimer, respectively, the total magnetization $m$ per elementary unit (all in $\mu_{\rm B}$ unit), and (b) the pair correlation functions $\langle\hat{\sigma}_{1,i}^{z}\hat{\sigma}_{2,i}^{z}\rangle$, $\langle\hat{S}_{1,i}^{x(y)}\hat{S}_{2,i}^{x(y)}\rangle$, $\langle\hat{S}_{1,i}^{z}\hat{S}_{2,i}^{z}\rangle$ as functions of the external magnetic field $B$ for the spin-$1/2$ Ising-Heisenberg orthogonal-dimer chain with the parameters $J_{H}/k_{\rm B} = 1.73$\,K, \mbox{$J_{I}^\prime/k_{\rm B} = 17.35$\,K}, $J_{I}/k_{\rm B} = 8.02$\,K, $g_{H} = 2.28$, $g_{I}= 18.54$ at the temperature $T = 0.05$\,K. The inset in~(a) shows the experimental magnetization data recorded for the polymer [Cu$_2$Dy$_2$]$_n$ in pulsed magnetic fields up to $10$\,T at the temperature $0.5$\,T and the respective theoretical prediction.}
	\label{fig:4}
\end{figure}

The various field-induced ground-state spin arrangements of the symmetric isotropic spin-$1/2$ Ising-Heisenberg orthogonal-dimer chain with the specific coupling constants \mbox{$J_{H}/k_{\rm B} = 1.73$\,K}, $J_{I}^\prime/k_{\rm B} = 17.35$\,K, $J_{I}/k_{\rm B} = 8.02$\,K and the gyromagnetic factors $g_{H} = 2.28$, $g_{I}= 18.54$ indicate a rich magnetization scenario of the heterobimetallic coordination polymer [Cu$_2$Dy$_2$]$_n$. The theoretical prediction of the isothermal magnetization curve $m(B)$ for [Cu$_2$Dy$_2$]$_n$ at a very low temperature $T = 0.05$\,K, based on the proposed quantum chain, together with the magnetic-field dependencies of the local Ising and Heisenberg sub-lattice magnetization $m_{I}$, $m_{H}$, which correspond to the Dy$^{3+}$\!-\,Dy$^{3+}$ and Cu$^{2+}$\!-\,Cu$^{2+}$ dimers, respectively, are depicted in Fig.~\ref{fig:4}(a). The correctness of the plotted magnetization curves is independently confirmed by the corresponding magnetic-field variations of the pair correlation functions $\langle\hat{\sigma}_{1,i}^{z}\hat{\sigma}_{2,i}^{z}\rangle$, $\langle\hat{S}_{1,i}^{z}\hat{S}_{2,i}^{z}\rangle$, $\langle\hat{S}_{1,i}^{x(y)}\hat{S}_{2,i}^{x(y)}\rangle$ shown in Fig.~\ref{fig:4}(b). Their low-temperature values immediately specify a type of spin correlations within dysprosium and cooper dimers in [Cu$_2$Dy$_2$]$_n$.

It can be seen from the Fig.~\ref{fig:4}(a) that the $m(B)$ curve exhibits a zero plateau and three intermediate plateaus at $9.27\mu_{\rm B}$, $16.26\mu_{\rm B}$, $18.54\mu_{\rm B}$ per elementary unit before reaching the saturation value $20.82\mu_{\rm B}$. The zero magnetization plateau can be detected only at very low magnetic fields $B<0.177$\,T. Both zero sub-lattice magnetization $m_{I} = m_{H} = 0$ in combination with the identical negative pair correlation functions $\langle\hat{\sigma}_{1,i}^{z}\hat{\sigma}_{2,i}^{z}\rangle = \langle\hat{S}_{1,i}^{z}\hat{S}_{2,i}^{z}\rangle = \langle\hat{S}_{1,i}^{x(y)}\hat{S}_{2,i}^{x(y)}\rangle = -0.25$ clearly prove that it relates to the quantum frustrated singlet phase $|0,0\rangle$ given by eigenvector~(\ref{eq:0,0}). On the other hand, the zero Heisenberg magnetization $m_{H} = 0$ and the half value of the Ising magnetization $m_{I} = 9.27\mu_{\rm B}$ indicate that the magnetization plateau at $9.27\mu_{\rm B}$, formed immediately behind the zero one in the magnetic-field range $(1.177-0.274)$\,T, arises due to polarization of exactly a half of the Dy$^{3+}$\!-\,Dy$^{3+}$ dimers into the magnetic-field direction. This very narrow finite plateau relates to the interesting quantum ferrimagnetic singlet-like phase $|1,0,0,0\rangle$ with two-fold broken translation symmetry [see Eq.~(\ref{eq:1,0,0,0})], as also evidenced by the corresponding low-temperature values of the pair correlation functions $\langle\hat{\sigma}_{1,i}^{z}\hat{\sigma}_{2,i}^{z}\rangle = 0$, $\langle\hat{S}_{1,i}^{z}\hat{S}_{2,i}^{z}\rangle = -0.25$, $\langle\hat{S}_{1,i}^{x(y)}\hat{S}_{2,i}^{x(y)}\rangle \approx -0.05$. The last two consecutive plateaus at the non-saturated values $16.26\mu_{\rm B}$ and $18.54\mu_{\rm B}$ of the total magnetization in the field intervals $(0.274-4.107)$\,T and $(4.107-6.366)$\,T reflect the field-induced stability of the phases with all Dy$^{3+}$\!-\,Dy$^{3+}$ dimers fully polarized into the magnetic-field direction, namely, the classical ferrimagnetic phase $|1,-1\rangle$ given by the eigenvector~(\ref{eq:1,-1}) and the quantum ferrimagnetic singlet phase $|1,0\rangle$ given by the eigenvector~(\ref{eq:1,0}), respectively. This statement is directly supported by the maximum value of the local Ising magnetization $m_{I} = 18.54\mu_{\rm B}$, the corresponding pair correlation function $\langle\hat{\sigma}_{1,i}^{z}\hat{\sigma}_{2,i}^{z}\rangle = 0.25$, and the following two sets of the Heisenberg sub-lattice magnetization and pair correlation functions: $m_{H} = -2.28\mu_{\rm B}$, $\langle\hat{S}_{1,i}^{z}\hat{S}_{2,i}^{z}\rangle = 0.25$, $\langle\hat{S}_{1,i}^{x(y)}\hat{S}_{2,i}^{x(y)}\rangle = 0$ and $m_{H} = 0$, $\langle\hat{S}_{1,i}^{z}\hat{S}_{2,i}^{z}\rangle = \langle\hat{S}_{1,i}^{x(y)}\hat{S}_{2,i}^{x(y)}\rangle = -0.25$.
The steep continuous changes in low-temperature magnetization curves and correlation functions, which can be observed in Fig.~\ref{fig:4} around the critical and saturation fields $B_{c1} = 0.177$\,T, $B_{c2} = 0.274$\,T, $B_{c3} = 4.107$\,T, $B_{sat} = 6.366$\,T clearly delimit stability regions for individual theoretically predicted ground-state arrangements of the polymeric compound [Cu$_2$Dy$_2$]$_n$. As can be expected, these changes are gradually smeared out by rising temperature until they completely disappear. In the case of total magnetization $m$, it happens so already at the relatively low temperature $T=0.5$\,K, which can be confirmed by not only the theoretical magnetization curve corresponding to the quantum mixed-spin orthogonal-dimer chain proposed in Sec.~\ref{sec:2}, but also the real experimental magnetization data recorded recorded for [Dy$_2$Cu$_2$]$_n$ [see the inset of Fig.~\ref{fig:4}(a)].

\subsection{Bipartite entanglement within {\rm Cu}$^{2+}$\!-\,{\rm Cu}$^{2+}$ dimers}
\label{subsec:3.3}

In the last subsection, we turn our attention to the bipartite entanglement between Cu$^{2+}$ magnetic ions forming the spin-$1/2$ dimers along the crystallographic $b$-axis of the coordination compound [Dy$_2$Cu$_2$]$_n$. The concurrence $C$ quantifying a degree of the entanglement within individual Cu$^{2+}$\!-\,Cu$^{2+}$ dimers is defined by Eq.~(\ref{eq:C}). To gain a complete picture, the obtained numerical results are discussed across the magnetization process as well as upon temperature increase.

\begin{figure*}[ht!]
	\centering
	\includegraphics[width=1.0\textwidth]{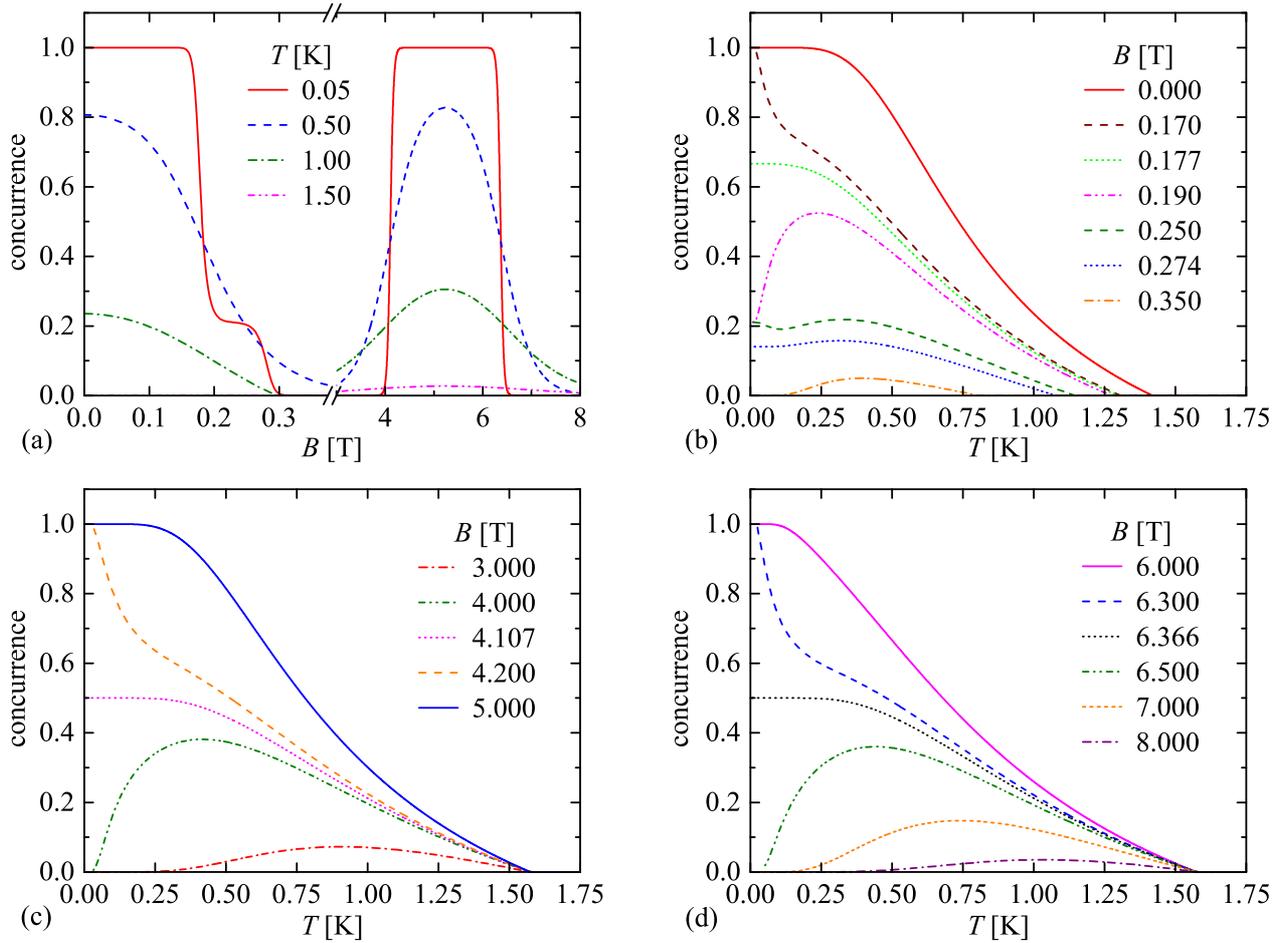}
	\vspace{-7mm}
	\caption{The concurrence $C$ reflecting the bipartite entanglement within cooper dimers of the compound [Dy$_2$Cu$_2$]$_n$ (a) as a function of the external magnetic field $B$ for several selected values of the temperature $T$ and (b)--(d) as a function of the temperature for various magnetic fields $B$. The results were calculated using the spin-$1/2$ Ising-Heisenberg orthogonal-dimer chain with the exchange interaction constants $J_{H}/k_{\rm B} = 1.73$\,K, $J_{I}^\prime/k_{\rm B} = 17.35$\,K, $J_{I}/k_{\rm B} = 8.02$\,K and the gyromagnetic factors $g_{H} = 2.28$, $g_{I}= 18.54$. 
	}
	\label{fig:5}
\end{figure*}

Figure~\ref{fig:5}(a) illustrates the isothermal dependencies of the concurrence $C$ on the external magnetic field $B$ for several selected values of the temperature $T$. The lowest-temperature one for $T=0.05$\,K (the solid red line) with two plateaus at  $C = 1$, one plateau at $C \approx 0.211$ and four abrupt changes near the critical and saturation fields $B_{c1} = 0.177$\,T, $B_{c2} = 0.274$\,T, $B_{c3} = 4.107$\,T, $B_{sat} = 6.366$\,T faithfully reflects a current strength of the quantum entanglement of horizontal cooper ionic pairs in individual ground states. The maximum concurrence $C = 1$ at very low magnetic fields up to $0.177$\,T and in the rather wide high-field range $(4.107-6.366)$\,T proves that the Cu$^{2+}$\!-\,Cu$^{2+}$ dimers are fully entangled only in the frustrated singlet phase $|0,0\rangle$ and the ferrimagnetic singlet phase $|1,0\rangle$. Moreover, the partial quantum entanglement within Cu$^{2+}$\!-\,Cu$^{2+}$ dimers can also be found in the ferrimagnetic singlet-like phase $|1,0,0,0\rangle$. The corresponding narrow plateau at non-trivial value of the concurrence $C \approx 0.211$ in the magnetic-field range $(0.177-0.274)$\,T indicates that it is roughly five times weaker than that in $|0,0\rangle$ and/or $|1,0\rangle$. The reason is the magnetic cooper dimers reside in the outstanding singlet-like states instead of the perfect singlet-dimer one [cf. the corresponding eigenvector~(\ref{eq:1,0,0,0}) with~(\ref{eq:0,0}) and~(\ref{eq:1,0})]. 
Besides, the low-temperature concurrence curve also contains two zero plateaus.  In agreement with the ground-state analysis, these occur at the relatively wide range of moderate magnetic fields $(0.274-4.107)$\,T and after exceeding the saturation field at $B>6.366$\,T, which confirms the classical character of the ferrimagnetic and saturation phases $|1,-1\rangle$ and $|1,1\rangle$, respectively.

It is apparent from Fig.~\ref{fig:5}(a) that an increase in temperature causes an evident decrease of the concurrence from its maximum value $C=1$ (weakening of the perfect bipartite entanglement between the Cu$^{2+}$ ionic pairs) above the both singlet ground states $|0,0\rangle$, $|1,0\rangle$. By contrast, a small rise of temperature invokes an increase of $C$ above the remaining phases $|1,0,0,0\rangle$, $|1,-1\rangle$, $|1,1\rangle$. This inverse trend is just temporary and can be ascribed to low-lying thermal excitations from the less entangled ground state $|1,0,0,0\rangle$ (the non-entangled phases $|1,-1\rangle$, $|1,1\rangle$) towards the more entangled one(s) $|0,0\rangle$ ($|1,0,0,0\rangle$ and/or $|1,0\rangle$). 

The above findings are demonstrated in more detail in Figs.~\ref{fig:5}(b)--(d), which show typical temperature variations of the concurrence $C$ for various magnetic fields $B$. In agreement with previous discussion, the quantity $C$ either monotonically decreases from its maximum zero-temperature asymptotic value with increasing temperature or shows a striking non-monotonous temperature dependence with a round maximum at moderate temperatures before it definitively falls to zero value at a certain threshold temperature. The former (standard) thermal trend can be observed if the quantum singlet phases $|0,0\rangle$, $|1,0\rangle$ constitute the ground state [see the curves corresponding to $B = 0$\,T,  $0.17$\,T in Fig.~\ref{fig:5}(b), $B = 4.2$\,T,  $5$\,T in Fig.~\ref{fig:5}(c), and $B = 6$\,T, $6.3$\,T in Fig.~\ref{fig:5}(d)], or if the applied magnetic field $B$ takes the specific values corresponding to the ground-state phase transitions $|0,0\rangle$\,--\,$|1,0,0,0\rangle$, \mbox{$|1,-1\rangle$\,--\,$|1,0\rangle$},  \mbox{$|1,0\rangle$\,--\,$|1,1\rangle$} [see the curves for $B_{c1} = 0.177$\,T in Fig.~\ref{fig:5}(b), $B_{c3} = 4.107$\,T in Fig.~\ref{fig:5}(c), and $B_{sat} = 6.366$\,T in Fig.~\ref{fig:5}(d)]. The unusual non-monotonous trend of $C$ can be observed if $B$ forces the system into the singlet-like state $|1,0,0,0\rangle$ or the completely non-entangled phases $|1,-1\rangle$, $|1,1\rangle$ at zero temperature [the curves corresponding to $B = 0.19$\,T,  $0.25$\,T, $0.35$\,T in Fig.~\ref{fig:5}(b), $B = 4$\,T, $3$\,T in Fig.~\ref{fig:5}(c), and $B = 6.5$\,T, $7$\,T, $8$\,T in Fig.~\ref{fig:5}(d)], as well as at the critical field $B_{c2} = 0.274$\,T corresponding to the ground-state phase transition $|1,0,0,0\rangle$\,--\,$|1,-1\rangle$ [see the dotted blue curve in Fig.~\ref{fig:5}(b)]. 

\section{Conclusion}
\label{sec:4}

The article presents the computational study of the ground-state spin arrangement and the bipartite entanglement between two exchange-coupled Cu$^{2+}$ magnetic ions forming the dimers along crystallographic $b$-axis of the 4f-3d heterometallic polymeric coordination compound [Dy$_2$Cu$_2$]$_n$. The study has been performed using the exactly solvable symmetric spin-$1/2$ Ising-Heisenberg orthogonal-dimer chain with the specific parameters $J_{H}/k_{\rm B} = 1.73$\,K, $J_{I}^\prime/k_{\rm B} = 17.35$\,K, $J_{I}/k_{\rm B} = 8.02$\,K,  $g_{H} = 2.28$, $g_{I}= 18.54$, whose combination recently provided quite satisfactory theoretical fit for known experimental magnetization data~\cite{Str20}.

The obtained numerical results point to five possible  ground-state spin arrangements of [Dy$_2$Cu$_2$]$_n$ upon gradual magnetic-field increase. Two of them are of non-entangled nature: the standard ferrimagnetic phase $|1,-1\rangle$ with the perfect antiparallel spin arrangement of the Dy$^{3+}$\!-\,Dy$^{3+}$ (Ising) and Cu$^{2+}$\!-\,Cu$^{2+}$ (Heisenberg) dimers and 
the fully saturated phase $|1,1\rangle$ with all spins polarized into the magnetic-field direction. In the other three ground-state spin arrangements, the Cu$^{2+}$\!-\,Cu$^{2+}$ dimers are quantum-mechanically entangled, whereas the orthogonal Dy$^{3+}$\!-\,Dy$^{3+}$ ones are either in two-fold degenerate antiferromagnetic states, if the frustrated singlet phase $|0,0\rangle$ is occurred, or fully polarized into the magnetic-field direction, if the ferrimagnetic singlet phase $|1,0\rangle$ is stable, or they regularly alternate between the ferro- and antiferromagnetic spin arrangement, if the ferrimagnetic singlet-like phase $|1,0,0,0\rangle$ with a two-fold breaking of translation symmetry constitutes  the ground state. The predicted sequence of the field-induced phase transitions $|0,0\rangle$\,--\,$|1,0,0,0\rangle$\,--\,$|1,-1\rangle$\,--\,$|1,0\rangle$\,--\,$|1,1\rangle$ has been independently identified in the low-temperature magnetization process of [Dy$_2$Cu$_2$]$_n$ as the series of zero plateau, three intermediate plateaus at the non-saturated values $9.27\mu_{\rm B}$, $16.26\mu_{\rm B}$, $18.54\mu_{\rm B}$ of the total magnetization per elementary unit and the high-field plateau at the saturation magnetization $20.82\mu_{\rm B}$, respectively.

The special attention has been paid to the investigation of the bipartite entanglement within Cu$^{2+}$\!-\,Cu$^{2+}$ dimers of [Dy$_2$Cu$_2$]$_n$. It has been demonstrated by the concurrence concept that the phenomenon is strongly magnetic-field dependent. Besides the low-field region $B<0.274$, where the frustrated singlet phase $|0,0\rangle$ with fully entangled Cu$^{2+}$\!-\,Cu$^{2+}$ dimers and the ferrimagnetic singlet-like phase $|1,0,0,0\rangle$ with partially entangled  Cu$^{2+}$\!-\,Cu$^{2+}$ dimers constitute the ground state, the Cu$^{2+}$ magnetic pairs forming the spin-$1/2$ dimers can be fully entangled into perfect singlet states also at within high-field region $B\in(4.107-6.366)$\,T due to stability of the phase $|1,0\rangle$. In general, the bipartite entanglement between two exchange-coupled Cu$^{2+}$ ions gradually weakens with increasing temperature until it completely vanishes. On the other hand, the non-monotonous temperature variations of the concurrence clearly point to the temperature-induced strengthening of the phenomenon above the partially entangled ground-state phase $|1,0,0,0\rangle$ and the thermal activation of the entangled states of the cooper dimers above the non-entangled ground states $|1,-1\rangle$ and $|1,1\rangle$.

As it was marginally indicated Sec.~\ref{sec:2}, the oximate \mbox{O--N} bridges between Cu$^{2+}$ and Dy$^{3+}$ ions are of two different strengths in real [Cu$_2$Dy$_2$Cu] cells~\cite{Oka09}. Therefore, the forthcoming theoretical examination of the magnetic properties of the polymeric coordination compound [Dy$_2$Cu$_2$]$_n$ will be performed through the more general asymmetric spin-$1/2$ Ising-Heisenberg orthogonal-dimer chain with two different inter-dimer exchange constants between Ising and Heisenberg spins. In addition to a closer approximation of the real system, the planned study may point to some deviations (shortcomings) in the use of the symmetric Ising-Heisenberg orthogonal-dimer chain discussed in the present article. 

\section*{Acknowledgment}
The work was funded by the Slovak Research and Development Agency under the contract No. APVV-20-0150 and by the Ministry of Education, Science, Research and Sport of the Slovak Republic under the grant No. VEGA 1/0301/20.


\end{document}